\documentclass[a4paper, conference, 10pt]{IEEEtran}
\usepackage{xcolor}
\usepackage{amssymb}
\usepackage{amsmath}
\usepackage[ruled,vlined]{algorithm2e}
\usepackage{cite}
\usepackage{mathtools}
\usepackage{array}
\usepackage{multicol}
\usepackage{multirow}
\usepackage{booktabs}
\usepackage{subfigure}

\newtheorem{proposition}{Proposition}[section]

\usepackage[top=1.90cm, bottom=4.29cm, left=1.29cm, right=1.29cm]{geometry}

\ifCLASSINFOpdf
\else
  \usepackage[dvips]{graphicx}
  \graphicspath{{Figures/eps/}}
\fi

\begin{document}

\title{Sharing is Caring: A Mobile Edge Computing Perspective}
\author{Nilanjan Biswas, Hamed Mirghasemi, Luc Vandendorpe\\
Universite Catholique de Louvain, Louvain la Neuve, Belgium\\
{(nilanjan.biswas,seyed.mirghasemi,luc.vandendorpe)}$@$uclouvain.be}

\maketitle

\begin{abstract}
In this paper, we consider a system model in conjunction with two major technologies in 5G communications, i.e., mobile edge computing and spectrum sharing. An IoT network, which does not have access to any licensed spectrum, carries its computational task offloading activities with help of spectrum sharing. The IoT network cooperates with a licensed spectrum holding network by relaying its data and in return gets access to the licensed spectrum. The licensed spectrum holding network focuses on throughput maximization, whereas, the IoT network tries to maximize task computation rate. We formulate an optimization problem giving importance to both networks’ interests. Typical IoT nodes may be energy-constrained, which is considered along with task computation time constraints. Parallel computation is considered while computing IoT nodes’ computational tasks at the mobile edge computing server, where the processor is allocated based on IoT nodes’ offloading capabilities. The advantage of such processor allocation is shown in the result section. Moreover, we also show how both IoT and licensed spectrum holding networks benefit from the spectrum sharing by caring for each other requirements, which echoes the fact, i.e., Sharing is Caring.
\end{abstract}


\IEEEpeerreviewmaketitle

\section{Introduction}
In modern wireless communication systems, we observe two important facts, i.e., a large number of wireless devices with huge data requirements and computation-intensive applications. According to Cisco [1], per capita will have 1.5 wireless devices and global mobile traffic is going to be as high as 77 exabytes/month. In that report, Cisco has also predicted the rise of computation-intensive tasks, e.g., facial recognition, augmented reality, virtual reality, highly interactive online gaming, internet of things (IoT), machine to machine (M2M) communications, etc. Two important questions arise under the above-mentioned scenarios: (a) How to accommodate the ever-increasing number of wireless devices and their data requirements within a limited spectrum? and (b) How to maintain latency requirement for computation-intensive tasks at energy and computation power-constrained wireless devices? Thanks to spectrum sharing and mobile edge computing (MEC) technologies, which help us in replying above queries. 

Both licensed \cite{hoglund2017overview} and unlicensed spectrum \cite{Nokia} may be shared at any/multiple domains, e.g., time, frequency, and space \cite{akyildiz2006next}. Typically two different techniques are followed for spectrum sharing, i.e., spectrum sensing \cite{biswas2018optimal} and cooperative relaying \cite{han2009cooperative}. We find the importance of spectrum sharing in 5G \cite{WinNT} and 6G \cite{tariq2019speculative} initiatives; more specifically, in large networks like IoT \cite{zhang2018spectrum}, where exclusive spectrum allocation may not be feasible. IoT nodes may operate in an unlicensed spectrum. However, due to uncontrolled interference, IoT nodes may not get satisfactory performance by using an unlicensed spectrum. In licensed spectrum, interference may be controlled by proper scheduling. But, for dense IoT networks (smart city), exclusive allotment of licensed spectrum for IoT networks may not be feasible. Therefore, for IoT networks, licensed spectrum sharing may be a useful idea. We find standardization effort on cellular spectrum sharing for IoT networks in \cite{hoglund2017overview}, where IoT nodes may be deployed inside the cellular spectrum.

Another important technology, which has gained attention for IoT networks is MEC \cite{hu2015mobile}. Typical IoT nodes are energy and computation power constrained, such that, their performances in terms of task computation may not be satisfactory. IoT nodes may take the help of a separate computing facility. In mobile edge computing (MEC), unlike cloud computing \cite{dinh2013survey}, a computing facility is installed at the edge of networks, which significantly reduces round trip delay. IoT nodes offload tasks to MEC servers following either partial or binary \cite{mao2017survey} offloading technique. In partial offloading, wireless devices may offload a fraction of computation tasks to MEC servers and the remaining are computed locally. In binary offloading, computation tasks are either offloaded to MEC servers or computed locally. Partial offloading may be preferred over binary offloading when computation tasks are granular (e.g., image processing) as it may help in reducing energy consumption at devices \cite{liang2019multiuser}, which is very important for IoT networks.


In this paper, like \cite{liu2019energy,biswas2020joint}, we consider cooperative relaying-based cellular spectrum sharing for an IoT network, where IoT nodes offload their computational tasks to a MEC server using the cellular spectrum. Following, we discuss the novelty of this work:
\begin{itemize}
	\item Both IoT and cellular networks participate in cooperative relaying. In Fig. 3(b) and 4(b), we show benefit for both IoT and cellular networks from the spectrum sharing, which echoes the fact, i.e., \textit{`Sharing is Caring'}. 
	\item Unlike \cite{liu2019energy}, we consider a joint optimization problem to solve relaying and computation parameters.
	\item Unlike \cite{liu2019energy,biswas2020joint}, we consider multiple offloading users and schedule them both for offloading and computation at the MEC server.
	\item We consider parallel computing and find optimal processor allocation based on offloading amount.
\end{itemize}
In Sections II, III, and IV, we discuss the system model, optimization problem, and some results, and conclude in Section V.

\section{System Model}
\begin{figure}[h!]
	\vspace{-4mm}
	\centering
	\includegraphics[trim=4cm 4.5cm 6cm 2cm,clip=true,width=5cm]{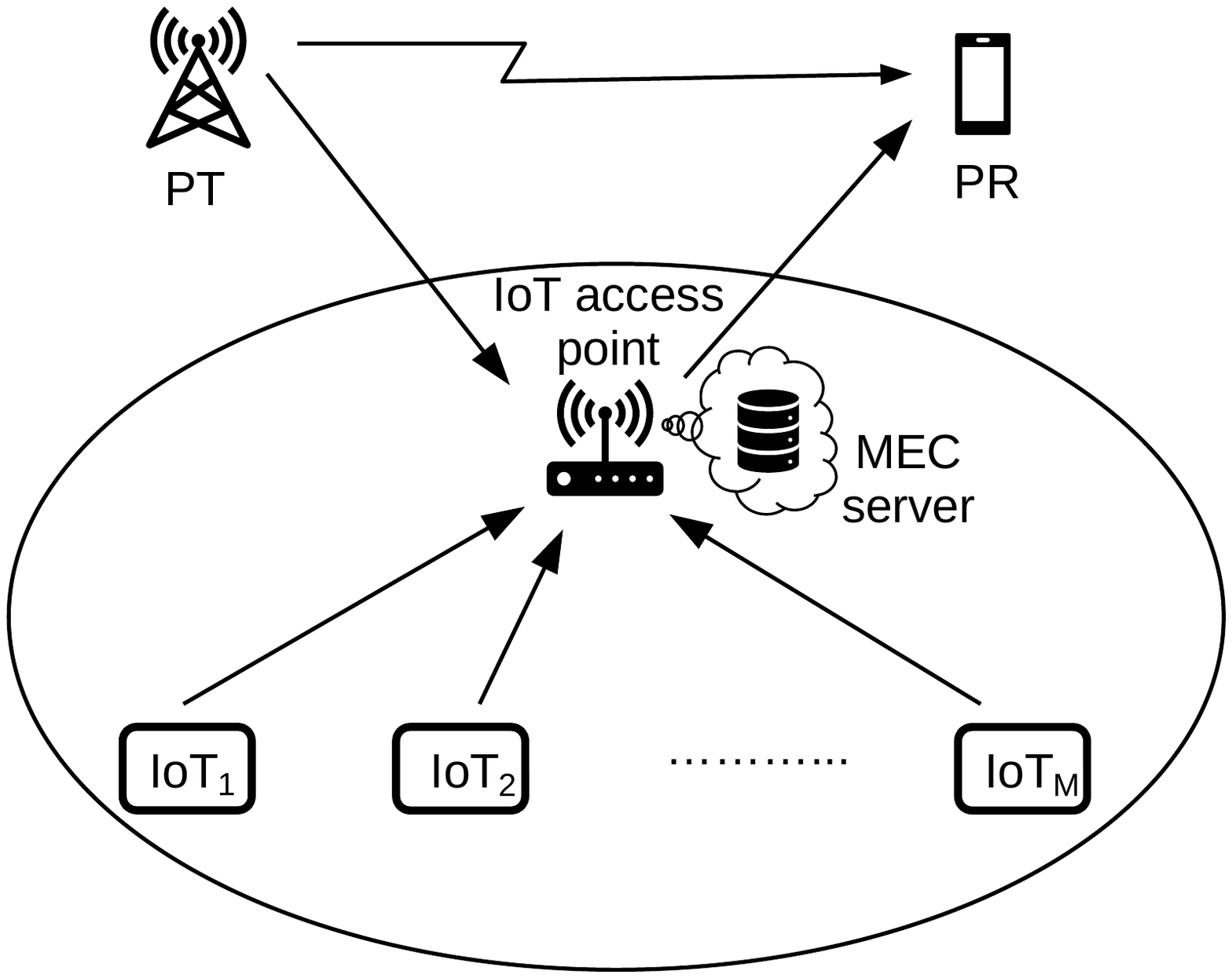}
	\DeclareGraphicsExtensions{.eps}
	\caption{Relaying based IoT network}
	\label{fig1}
	\vspace{-4mm}
\end{figure}

In Fig.~\ref{fig1}, we show the logical diagram of our system model, where we consider a cellular network and an IoT network. Following the state of the art, we denote the cellular network by primary network. We consider a single transmitter and its corresponding receiver for the cellular network, which are denoted by the primary transmitter (PT) and the primary receiver (PR), respectively \footnote{We have considered single PT and PR to keep the system model simple and to convey the idea of this work. Our future plan includes scaling up the network with multiple users.}. There is a IoT network with IoT nodes, i.e., $\left\{\text{IoT}_k\right\}_{k=1}^M$, and a IoT access point integrated with a MEC server. The IoT network does not have access to the cellular spectrum. Due to poor channel gain between the PT and the PR, the received rate at the PR may be poor. On the other hand, due to computation power and energy constraints, IoT nodes' computation rates may be poor. Under such a scenario, we consider cooperation among two networks, such that, users of these networks can satisfy their requirements. The IoT access point helps the PT in relaying its data to the PR, such that, the PR gets its required rate. It is to be noted that cooperative relaying is already there for LTE-Advanced \cite{rel11}, which may increase LTE networks' coverage. As remuneration, the primary network allows the IoT network to access the cellular spectrum. IoT nodes may perform partial offloading to send their computational tasks to the MEC server at the IoT access point over the cellular spectrum. Due to two separate hardware blocks for transceiver and computation, an IoT node can offload and locally compute computational tasks simultaneously.	
We consider following assumptions:

a) Both primary and IoT networks follow synchronized time slot structure.

b) During a slot duration, users for both networks are static, such that, we can consider a slot duration below coherence time. Therefore,  channel gains remain same during a time slot.

c) IoT nodes' computational tasks are finely granular \cite{biswas2020joint}, such that, tasks can be partitioned into bits \footnote{Please note that tasks can be partitioned into subset of any size. Hence, bit level partitioning may serve as an upper bound on performance.}.
  
\begin{figure}[h!]
	\vspace{-4mm}
	\centering
	\includegraphics[trim=4cm 8cm 4cm 6.5cm,clip=true,width=6cm]{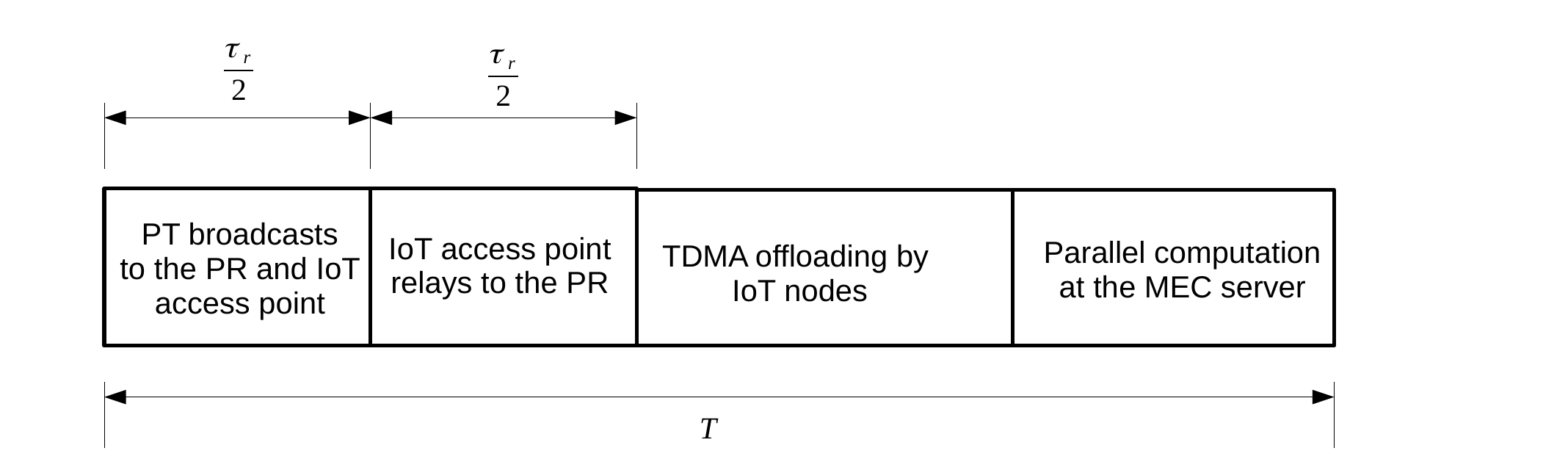}
	\DeclareGraphicsExtensions{.eps}
	\caption{Different phases based on communication and computation}
	\label{fig2}
	\vspace{-4mm}
\end{figure}

During each slot, the time frame duration is considered to be $T$.  Based on activities at the PT, PR, IoT nodes, and IoT access point transceivers, we divide a frame duration into different phases, which is shown in Fig.~\ref{fig2}. At the beginning of a frame, the PT broadcasts data to the PR and the IoT access point for $\tau_r/2$ duration. After receiving the PT's data, the IoT access point relays it to the PR for $\tau_r/2$ duration.\footnote{Different time durations for broadcasting and relaying might be considered, which brings more variables in the optimization problem. We leave this for future work.} Once the relaying is over, IoT nodes offload their computational tasks to the MEC server at the IoT access point following time divisional multiple access (TDMA) protocol. Like \cite{liang2019multiuser}, we consider that the MEC server starts parallel processing after receiving all IoT nodes computational tasks \footnote{This kind of strategy has been adopted in \cite{liang2019multiuser}. A different approach may be considered, where task reception at the IoT access point and computation at the MEC server go simultaneously, which will make the scenario more general and we leave that as our future work.}. The MEC server computes IoT nodes' computational tasks and feedbacks them to IoT nodes. Under the assumption of the IoT access point's high transmission power and lower number of feedback bits, we neglect downloading time durations at IoT nodes like \cite{liu2019energy}. Note that, IoT nodes may perform local computing from the very beginning of a frame, which does not hamper the offloading process due to separate hardware blocks for computation and communication. 

\textit{Possible application:} Deployment of IoT nodes in LTE spectrum has already been considered in \cite{hoglund2017overview}. Moreover, cooperative relaying is also there in LTE networks \cite{rel11}. Therefore, our system model may be implemented in LTE networks.  

In the following subsections, we discuss different metrics for both networks, which we maximize in this paper. More specifically, we discuss relaying and computation rates for the primary and the IoT networks, respectively. We also define constraints, which make the problem more realistic.
\subsection{Relaying rate of the primary network}
We consider DF relaying in this work. Following \cite{vandendorpe2008power}, we can write the PT's rate as:
\begin{equation}
	R_{\text{relay}} = \frac{\tau_r B_w}{2}\log_2\left\{1+\min(\gamma_{\text{PT-AP}},\gamma_{\text{PT-PR+AP-PR}})\right\} 
	\label{eq1}
\end{equation}  
where $B_w$ is the bandwidth of the cellular spectrum and $
\gamma_{\text{PT-PR+AP-PR}} = \frac{P_{\text{PT}}\left|h_{\text{PT-PR}}\right|^2d_{\text{PT-PR}}^{-\eta_1}}{\sigma^2}+
\frac{P_{\text{AP}}\left|h_{\text{AP-PR}}\right|^2d_{\text{AP-PR}}^{-\eta_2}}{\sigma^2},
\gamma_{\text{PT-AP}} = \frac{P_{\text{PT}}\left|h_{\text{PT-AP}}\right|^2d_{\text{PT-AP}}^{-\eta_3}}{\sigma^2}$,
 $P_{\text{PT}}$ and $P_{\text{AP}}$ are the PT's transmission and the IoT access point's relaying powers, respectively, $\sigma^2$ denotes total noise power at the receiver (we consider identical noise power at all receivers). Distances between different entities are denoted by $d_{x-y}$ and channel gains are denoted by $h_{x-y}, x,y\in \left\{\text{PT,PR,IoT access point}\right\}$. We consider block fading, such that, channel gains remain same throughout a slot duration. $\eta_{1-3}$ are path loss exponents.

\subsection{Computation rates for the IoT network}
In the IoT network, IoT nodes may perform both local computation and partial offloading for computation at the MEC server. Following, we discuss those two rates.

\subsubsection{Local computation rate}
We characterize the computation task of IoT$_k$ by a positive parameter tuple $\left\langle C_{k},f_{k}\right\rangle$, where $C_{k}$ is the number of CPU cycles required to compute a bit at IoT$_k$ and $f_{k}$ is IoT$_k$'s CPU frequency (CPU cycles/second). IoT$_k$ can compute $R_{k}$ bits for the duration $t_{\text{loc},k}$, where:
\begin{equation}
R_{k}=\frac{t_{\text{loc},k}f_{k}}{C_{k}}.
\label{eq2-1}
\end{equation} 

\subsubsection{Offloading rate}
IoT$_k,\forall{k}$, may send a partial amount of computational tasks to the MEC server at the IoT access point. The IoT access point allocates time durations among IoT nodes for offloading their computational tasks. IoT$_k$ can offload $R_{\text{MEC},k}$ bits, which has following upper bound:
\begin{equation}
R_{\text{MEC},k}\leq t_{\text{off},k}B_w\log_2\left(1+\frac{e_{\text{off},k}|h_{k}|^2d_{k}^{-\eta_4}}{t_{\text{off},k}\sigma^2}\right), 
\label{eq2-2}
\end{equation} 
where $t_{\text{off},k},e_{\text{off},k}$,  $h_{k}$, $d_{k}$, $\eta_4$ are IoT$_k$'s offloading time duration, energy, channel gain, distance, and path loss exponent to the IoT access point. 

\subsection{Different constraints}
We consider constraints for both networks, which take care of benefits and limitations for users in both networks.

\subsubsection{Guaranteed relaying rate}
The primary network allows the IoT network to access the cellular spectrum only when the effective rate at the PR is guaranteed above the rate of the direct link between the PT and PR, i.e.,
\begin{equation}
R_{\text{relay}} \geq TB_w\log_2(1+\frac{P_{\text{PT}}\left|h_{\text{PT-PR}}\right|^2d_{\text{PT-PR}}^{-\eta_1}}{\sigma^2}). 
\label{eq2-4}
\end{equation} 

\subsubsection{Frame duration constraint}
Both local computations at IoT nodes and remote computation at the MEC server should be completed within the frame duration, i.e., $T$. As IoT$_k, \forall{k},$ has separate hardware blocks for local computation and wireless transmissions/receptions, we consider two separate constraints for local computation and remote computation as given in Equation~\eqref{eq2-5} and \eqref{eq2-6}, respectively, $\forall{k}$:
\begin{subequations}
	\begin{align}
		t_{\text{loc},k} &\leq T, \label{eq2-5}\\
		\frac{R_{\text{MEC},k}C_{k}}{f_{\text{MEC},k}} &\leq T-\tau_r-\sum_{k=1}^M t_{\text{off},k} \label{eq2-6},
	\end{align}
\end{subequations} 
where $f_{\text{MEC},k}$ is allotted CPU frequency (CPU cycles/second) for IoT$_k$ at the MEC server. In Equation~\eqref{eq2-6}, we show that the remote computation duration for IoT$_k$ should be less than the remaining time after relaying and total offloading as has been shown in Fig.~\ref{fig2}.

\subsubsection{Limitation of energy at IoT nodes}
We consider IoT$_k, \forall{k},$ having finite energy at its battery to complete computations and communications, i.e.,
\begin{equation}
\frac{w_{k}t_{\text{loc},k}f_{k}^3}{C_{k}}+e_{\text{off},k} \leq E_k, 
\label{eq2-7}
\end{equation} 
where $w_{k}$ is coefficient depending on chip architecture.

\subsubsection{Maximum CPU frequency at the MEC server} 
We assume the MEC server can afford maximum $f_{\text{MEC}}^\text{max} $ CPU cycles/second during computing IoT nodes' tasks. The MEC server divides the computation power among multiple IoT nodes and parallelly compute their tasks. Therefore, if IoT$_k$ is allotted $f_{\text{MEC},k}$ CPU cycles/second, then we can write:
\begin{equation}
\sum_{k=1}^M f_{\text{MEC},k} \leq f_{\text{MEC}}^\text{max}, 
\label{eq2-9}
\end{equation}

\section{System optimization}
In our system model, there are two different networks, which have two different interests. For the primary network, the PT will try to maximize the relaying rate; whereas, in the IoT network, the computation rate is of interest. We consider a single weighted objective function which considers both networks' interests, and form following optimization problem:
\begin{subequations}
	\begin{align}
	P1: &\underset{\tau_r,\mathbf{t_{\text{loc}}},\mathbf{t_{\text{off}}},\mathbf{e_{\text{off}}},\mathbf{R_{\text{MEC}}},\mathbf{f_{\text{MEC}}}}{\text{maximize}}\hspace{1mm}  \alpha R_{\text{relay}}+(1-\alpha)\sum_{k=1}^M[R_{k}+R_{\text{MEC},k}] \label{obj}\\
	&\text{subject to:}\hspace{5mm} \text{Equation}~\eqref{eq2-2}, \eqref{eq2-4}, \eqref{eq2-5}, \eqref{eq2-6}, \eqref{eq2-7},  \eqref{eq2-9},
	\end{align}
\end{subequations}
where $\alpha\in \left[0,1\right]$, $\mathbf{R_{\text{MEC}}}=\left\{R_{\text{MEC},k}\right\}_{k=1}^M$, $\mathbf{t_{\text{loc}}}=\left\{t_{\text{loc},k}\right\}_{k=1}^M$, $\mathbf{e_{\text{off}}}=\left\{e_{\text{off},k}\right\}_{k=1}^M$, $\mathbf{t_{\text{off}}}=\left\{t_{\text{off},k}\right\}_{k=1}^M$, and $\mathbf{f_{\text{MEC}}}=\left\{f_{\text{MEC},k}\right\}_{k=1}^M$.

We observe that the optimization problem $P1$ is non-convex due to division of $R_{\text{MEC},k}$ and $f_{\text{MEC},k}$ in Equation~\eqref{eq2-6}. We analyse optimal CPU allocation and present our findings in following proposition:
\begin{proposition}\label{prop1}
For $C_{k}=C, \forall{k}$, following statements are optimal:
\begin{itemize}
		\item Both Equation~\eqref{eq2-6} and \eqref{eq2-9} hold at equality
		\item Optimal allotted CPU frequency to IoT$_k$ is:
		\begin{align}
		f_{\text{MEC},k} = f_{\text{MEC}}^\text{max}\frac{R_{\text{MEC},k}}{\sum_{k=1}^M R_{\text{MEC},k}}
		\end{align}
\end{itemize}
\begin{proof}
We prove the first statement of this proposition by law of contradiction. Let us assume that optimal values for the optimization problem $P1$ are $\tau_r^*,\left\{t_{\text{off},k}^*\right\}_{k=1}^M,\left\{R_{\text{MEC},k}^*\right\}_{k=1}^M$, and $\left\{f_{\text{MEC},k}^*\right\}_{k=1}^M$, such that, either or both Equation~\eqref{eq2-6} and \eqref{eq2-9} hold at inequality. From Equation~\eqref{obj}, we can observe that objective function increases with offloaded bits from IoT nodes. However, offloaded bits should be computed within the frame duration as has been shown in Equation~\eqref{eq2-6}. Computation time for IoT$_k$ will reduce as the value for $f_k$ is increased or in other way, we can say that the MEC server can compute more for IoT$_k$ for higher values of $f_k$. From this logic, we can say that in optimal case, Equation~\eqref{eq2-9} holds equality. Now, we prove that Equation~\eqref{eq2-6} holds equality for optimal case. We can prove that using two facts, i.e., relaying rate increases for increasing $\tau_r$ and computation rate increases with $t_{\text{off},k}, \forall{k}$. From the structure of $P1$, it can be observed that both of these facts help us in increasing the value for objective function while satisfying constraints. Hence, we can increase value for $\tau_r$ and $t_{\text{off},k}$ to get equality in Equation~\eqref{eq2-6}.

Now, we prove the second statement of this proposition. As Equation~\eqref{eq2-6} holds at equality, for $C_k=C, \forall{k},$ in optimal case, we get:
\begin{align}
	\frac{R_{\text{MEC},1}}{f_{\text{MEC},1}} =..= \frac{R_{\text{MEC},k}}{f_{\text{MEC},k}}=...=\frac{R_{\text{MEC},M}}{f_{\text{MEC},M}}.
\end{align} 
Using basic algebraic rule, we get following relation:
\begin{align}
	\frac{R_{\text{MEC},1}}{f_{\text{MEC},1}} =..= \frac{R_{\text{MEC},M}}{f_{\text{MEC},M}}=\frac{\sum_{k=1}^M R_{\text{MEC},k}}{\sum_{k=1}^M f_{\text{MEC},k}}.
\end{align}
From above relation and equality in Equation~\eqref{eq2-9}, we get 
 $f_{\text{MEC},k} = f_{\text{MEC}}^\text{max}\frac{R_{\text{MEC},k}}{\sum_{k=1}^M R_{\text{MEC},k}}$.
\end{proof}	
\end{proposition}	

In our further discussion, we consider $C_{k}=C$, which is valid when IoT nodes' tasks are identical. Following Proposition~\ref{prop1}, we rewrite $P1$ as:
\begin{subequations}
	\begin{align}
	P2: &\underset{\tau_r,\mathbf{t_{\text{loc}}},\mathbf{t_{\text{off}}},\mathbf{e_{\text{off}}},\mathbf{R_{\text{MEC}}}}{\text{maximize}}\hspace{2mm}  \alpha R_{\text{relay}}+(1-\alpha)\sum_{k=1}^M[R_{k}+R_{\text{MEC},k}] \nonumber\\
	&\text{subject to:}\hspace{5mm} \text{Equation}~\eqref{eq2-2}, \eqref{eq2-4}, \eqref{eq2-5}, \eqref{eq2-7},  \nonumber\\
	& \frac{C}{f_{\text{MEC}}^{\text{max}}}\sum_{k=1}^M R_{\text{MEC},k} = T-\tau_r-\sum_{k=1}^M t_{\text{off},k}. \label{P5_cons1}
	\end{align}
\end{subequations} 

\subsection{Feasibility condition}
We observe that for the constraint on relaying rate as given in Equation~\eqref{eq2-4}, there will be a condition on $\tau_r$, which makes the optimization problem $P2$ feasible, i.e., $
\tau_r^{\text{lb}} \leq 1,$ where  $\tau_r^{\text{lb}}=\frac{2\log_2(1+\frac{P_{\text{PT}}\left|h_{\text{PT-PR}}\right|^2d_{\text{PT-PR}}^{-\eta_1}}{\sigma^2})}{\log_2\left\{1+\min(\gamma_{\text{PT-MEC}},\gamma_{\text{PT-PR+MEC-PR}})\right\}}$, is the lower bound on $\tau_r$. In following subsection, we show the solution method when the optimization problem $P2$ satisfies the feasibility condition.

\subsection{Solution method}
Corresponding dual function for $P2$ becomes:
\begin{align}
g(\zeta,\boldsymbol{\mu},\boldsymbol{\lambda},\boldsymbol{\eta})=&\underset{\tau_r,\mathbf{t_{\text{loc}}},\mathbf{t_{\text{off}}},\mathbf{e_{\text{off}}},\mathbf{R_{\text{MEC}}}}{\text{maximize}}\hspace{2mm}  L \nonumber\\
&\text{subject to:}\hspace{10mm} \text{Equation}~\eqref{eq2-4}.
\label{decomp}
\end{align} 
where 
\begin{align}
L &= \alpha R_{\text{relay}}+(1-\alpha)\sum_{k=1}^M[R_{k}+R_{\text{MEC},k}]
	-\nonumber\\
	&\zeta(\tau_r+\sum_{k=1}^M t_{\text{off},k}+\sum_{k=1}^M\frac{R_{\text{MEC},k}C}{f_{\text{MEC}}^\text{max}} - T)-\sum_{k=1}^M \mu_k(t_{\text{loc},k}
	-T)\nonumber\\
	&-\sum_{k=1}^M \lambda_k\{\frac{w_{k}t_{\text{loc},k}f_{k}^3}{C}+e_{\text{off},k} - E_k\}- \nonumber\\
	&\sum_{k=1}^M \eta_k\left\{R_{\text{MEC},k}- t_{\text{off},k}B_w\log_2\left(1+\frac{e_{\text{off},k}|h_{k}|^2d_{k}^{-\eta_4}}{t_{\text{off},k}\sigma^2}\right)\right\};
\end{align}	
$\zeta,\boldsymbol{\mu}=\{\mu_k\}_{k=1}^M,$  $\boldsymbol{\lambda}=\{\lambda_k\}_{k=1}^M,$ and $\boldsymbol{\eta}=\{\eta_k\}_{k=1}^M$ are Lagrange multipliers for Equation~\eqref{P5_cons1}, \eqref{eq2-5},  \eqref{eq2-7}, and \eqref{eq2-2}, respectively.

From $P2$, it can be observed that the optimization problem is convex. Therefore, we can get the optimal solution by solving following dual problem as duality gap is zero \cite{boyd2004convex}:
\begin{subequations}
	\begin{align}
	D1:\hspace{2mm} &\underset{\zeta,\boldsymbol{\mu},\boldsymbol{\lambda},\boldsymbol{\eta}}{\text{maximize}}\hspace{2mm}  g(\zeta,\boldsymbol{\mu},\boldsymbol{\lambda},\boldsymbol{\eta}) \\
	&\text{subject to:}\hspace{5mm} \zeta\geq 0, \mu_k\geq 0, \lambda_k \geq 0,\eta_k \geq 0, \forall{k}.
	\end{align}
\end{subequations}
We observe that for given $\zeta$, we can decompose the problem as given in Equation~\eqref{decomp}, into two sub-problems:
\begin{subequations}
	\begin{align}
		&\underset{\tau_r}{\text{maximize}}\hspace{2mm} \alpha R_{\text{relay}}-\zeta \tau_r, \hspace{2mm} \text{subject to:} \hspace{2mm} \text{Equation}~\eqref{eq2-4}\\
		&\underset{\mathbf{t_{\text{loc}}},\mathbf{t_{\text{off}}},\mathbf{e_{\text{off}}},\mathbf{R_{\text{MEC}}}}{\text{maximize}} L-\alpha R_{\text{relay}}+\zeta \tau_r 
		,\text{subject to:} \text{Equation}~\eqref{eq2-4} \label{cons56} 
	\end{align}
\end{subequations}
From Karush-Kuhn-Tucker's conditions, we can write following proposition. Due to brevity, we have omitted detailed discussion on it.
\begin{proposition}\label{prop3}
	For given $\zeta$, $\mu_k, \lambda_k$, and $\eta_k$, optimal $t_{\text{off},k}, R_{\text{MEC},k}, t_{\text{loc},k}, \forall{k}$ and $\tau_r$ are:
	\small{
	\begin{subequations}
	\begin{align}
		e_{\text{off},k} &= \rho_k t_{\text{off},k} \label{eqq1}\\
		t_{\text{off},k} &=\left\{ \begin{array}{rl}	&T-\tau_r^{\text{lb}}, \hspace{20mm}d_{k}<0\\
		&\left[0,T-\tau_r^{\text{lb}}\right], \hspace{14mm}d_{k}=0\\
		&0, \hspace{29mm}d_{k}>0
		\end{array} \right.\label{eqq2}\\
		t_{\text{loc},k} &= \left\{ \begin{array}{rl}	&T, \hspace{27mm}u_{k}<0\\
		&\left[0,T\right], \hspace{22mm}u_{k}=0\\
		&0, \hspace{29mm}u_{k}>0
		\end{array} \right.\label{eqq3}\\
		R_{\text{MEC},k} &=\left\{ \begin{array}{rl}	&t_{\text{off},k}B_w\log_2\left(1+\rho_k|h_{k}|^2d_{k}^{-\eta_4}/\sigma^2\right), \hspace{2mm}v_{k}<0\\
		&\left[0,t_{\text{off},k}B_w\log_2\left(1+\rho_k|h_{k}|^2d_{k}^{-\eta_4}/\sigma^2\right)\right], \hspace{2mm}v_{k}=0\\
		&0, \hspace{2mm}v_{k}>0
		\end{array} \right.\label{eqq5}\\	
		\tau_r &=\left\{ \begin{array}{rl}	&T, \hspace{20mm}t<0\\
		&\left[\tau_r^{\text{lb}},T\right], \hspace{12mm}t=0\\
		&\tau_r^{\text{lb}}, \hspace{19mm}t>0
		\end{array} \right.\label{eqq4}
	\end{align}
	\end{subequations}}\normalsize
where $\rho_k=\left[\frac{\eta_kB_w}{\lambda_k\ln2}-\frac{\sigma^2}{|h_{k}|^2d_{k}^{-\eta_4}}\right]^+, [x]^+=\max\left\{0,x\right\}, u_{k}=\mu_k+\lambda_kw_kf_k^3/C-(1-\alpha)f_k/C, d_k= \zeta-\eta_kB_w\log_2\left(1+\frac{\rho_k|h_{k}|^2d_{k}^{-\eta_4}}{\sigma^2}\right)+\eta_kB_w\log_2(e)\frac{\rho_k|h_{k}|^2d_{k}^{-\eta_4}}{\rho_k|h_{k}|^2d_{k}^{-\eta_4}+\sigma^2}$, $v_k = \zeta\frac{C}{f_{\text{MEC}}^{\text{max}}}+\eta_k-(1-\alpha)$, $t=\zeta-\frac{\alpha B_w}{2}\log_2\left\{1+\min(\gamma_{\text{PT-AP}},\gamma_{\text{PT-PR+AP-PR}})\right\}$.
\end{proposition} 

Once we get optimal values for primal variables, i.e., $\tau_r,\mathbf{t_{\text{loc}}},\mathbf{t_{\text{off}}},\mathbf{e_{\text{off}}},\mathbf{R_{\text{MEC}}}$, from Proposition~\ref{prop3}, we can find optimal dual variables, i.e., $(\zeta^*,\boldsymbol{\mu}^*,\boldsymbol{\lambda}^*,\boldsymbol{\eta}^*)$. We use subgradient based method to find optimal $(\zeta^*,\boldsymbol{\mu}^*,\boldsymbol{\lambda}^*,\boldsymbol{\eta}^*)$, where subgradients for $(\zeta,\mu_k,\lambda_k,\eta_k)$ are: $(\tau_r+\sum_{k=1}^M t_{\text{off},k}+\sum_{k=1}^M\frac{R_{\text{MEC},k}C}{f_{\text{MEC}}^\text{max}} - T,t_{\text{loc},k}-T,\frac{w_{k}t_{\text{loc},k}f_{k}^3}{C}+e_{\text{off},k} - E_k,R_{\text{MEC},k}- t_{\text{off},k}B_w\log_2\left(1+\frac{\rho_k|h_{k}|^2d_{k}^{-\eta_4}}{\sigma^2}\right))$. With optimal $\eta_k^*$ and $\lambda_k^*$, we find $\rho_k^*$ from Equation~\eqref{eqq1}, $\forall{k},$ and then solve following linear programming:
\begin{subequations}
	\begin{align}
	P3: &\underset{\tau_r,\mathbf{t_{\text{loc}}},\mathbf{t_{\text{off}}},\mathbf{R_{\text{MEC}}}}{\text{maximize}}\hspace{2mm}  \alpha R_{\text{relay}}+(1-\alpha)\sum_{k=1}^M[R_{k}+R_{\text{MEC},k}] \nonumber\\
	&\text{subject to:}\hspace{5mm} \text{Equation}~\eqref{eq2-4}, \eqref{eq2-5}, \eqref{P5_cons1}, \eqref{eq2-7},\nonumber\\
	& R_{\text{MEC},k}\leq t_{\text{off},k}B_w\log_2\left(1+\frac{\rho_k^*|h_{k}|^2d_{k}^{-\eta_4}}{\sigma^2}\right). \label{P4_cons1}
	\end{align}
\end{subequations} 

In following algorithm, we zest solution steps to solve the optimization problem $P2$: 
\begin{algorithm}[H]
	\scriptsize{
	\nl Initialize: $\zeta,\boldsymbol{\mu}, \boldsymbol{\lambda}, \boldsymbol{\eta}$\;	
	\nl \Repeat{Until $\zeta, \boldsymbol{\mu}, \boldsymbol{\lambda}$, $\boldsymbol{\eta}$ converge}{
		\nl Evaluate $R_{\text{MEC},k},t_{\text{off},k}, e_{\text{off},k}, t_{\text{loc},k}, \forall{k}$ and $\tau_r$ from Proposition~\ref{prop3}\;
		\nl Calculate subgradients for $\zeta, \boldsymbol{\mu}, \boldsymbol{\lambda},\boldsymbol{\eta}$ and update using ellipsoid method\;
	}
	\nl \KwResult{Obtain optimal ratio, i.e., $\rho_k^*$ and then solve linear programming of the optimization problem $P3$ to compute optimal $R_{\text{MEC},k}^*, t_{\text{off},k}^*, t_{\text{loc},k}^*, \forall{k}$ and $\tau_r^*$}
	\caption{Optimal solution for $P2$\label{algo1}}}
\end{algorithm}

\section{Results and discussions}
In Table~\ref{tab1}, we show different parameters' values, under which we have generated our results. As IoT nodes may be deployed in LTE spectrum \cite{hoglund2017overview}, we have considered the spectrum bandwidth according to LTE spectrum bandwidth.  
\begin{table}[h!]
	\centering
	\caption{Different parameters' values\label{tab1}}
	\begin{tabular}{|l|l|l|l|l}
		\cline{1-4}
		Parameters & Values      & Parameters  & Values     &  \\ \cline{1-4}
		$T$        & 0.1 sec.    & $w_{k}$    & $10^{-28}$ &  \\ \cline{1-4}
		$B_w$      & 1.4 MHz.     & $d_{\text{PT-PR}}$    & $100$ meters  &  \\ \cline{1-4}
		$\sigma^2$      & -132.24 dBm & $f_{k}$   & $10^{10}$ Hz.&  \\ \cline{1-4}
		$P_{PT}$   & 43 dBm      & $C_{k},\forall{k}$   & $10^4$ &  \\ \cline{1-4}
		$P_{\text{AP}}$  & 30 dBm      & $f_{\text{MEC}}$   & $10^{12}$ Hz. &  \\ \cline{1-4}
	\end{tabular}
\end{table}
We consider twenty IoT nodes in the IoT network. Among other parameters, we consider channels between different entities are Rayleigh faded, such that, their channel gains follow exponential distributions. $|h_{\text{AP-PR}}|^2,|h_{\text{PT-PR}}|^2$. $|h_{\text{PT-AP}}|^2$ are exponentially distributed with mean values 1, $10^-3$ \footnote{We consider lower channel gain between the PT and PR, which is realistic for cooperative relaying.}, and 1, respectively; whereas, $|h_{k}|^2, \forall{\text{IoT}_k}$, follow exponential distribution with mean value 5. Moreover, we consider the IoT access point is on the straight line joining the PT and PR, such that, $d_{\text{AP-PR}}=d_{\text{PT-PR}}-d_{\text{PT-AP}}$. We consider $d_{k}=10$ meters, $\forall{\text{IoT}_k}$. Different path loss exponents are $\eta_{1-3}=4$ and $\eta_4=2$. In order to understand cooperation benefits from the spectrum sharing, we evaluate gains for both primary and IoT networks and plot them in Fig.~\ref{res_fig1}(b) and \ref{res_fig2}(b), respectively.

\subsection{Primary network's relaying rate and gain for varying $\alpha$}
\begin{figure}[!h]
	\vspace{-2mm}
	\centering
	\includegraphics[trim=2cm 10cm 2cm 8cm,clip=true,width=8cm]{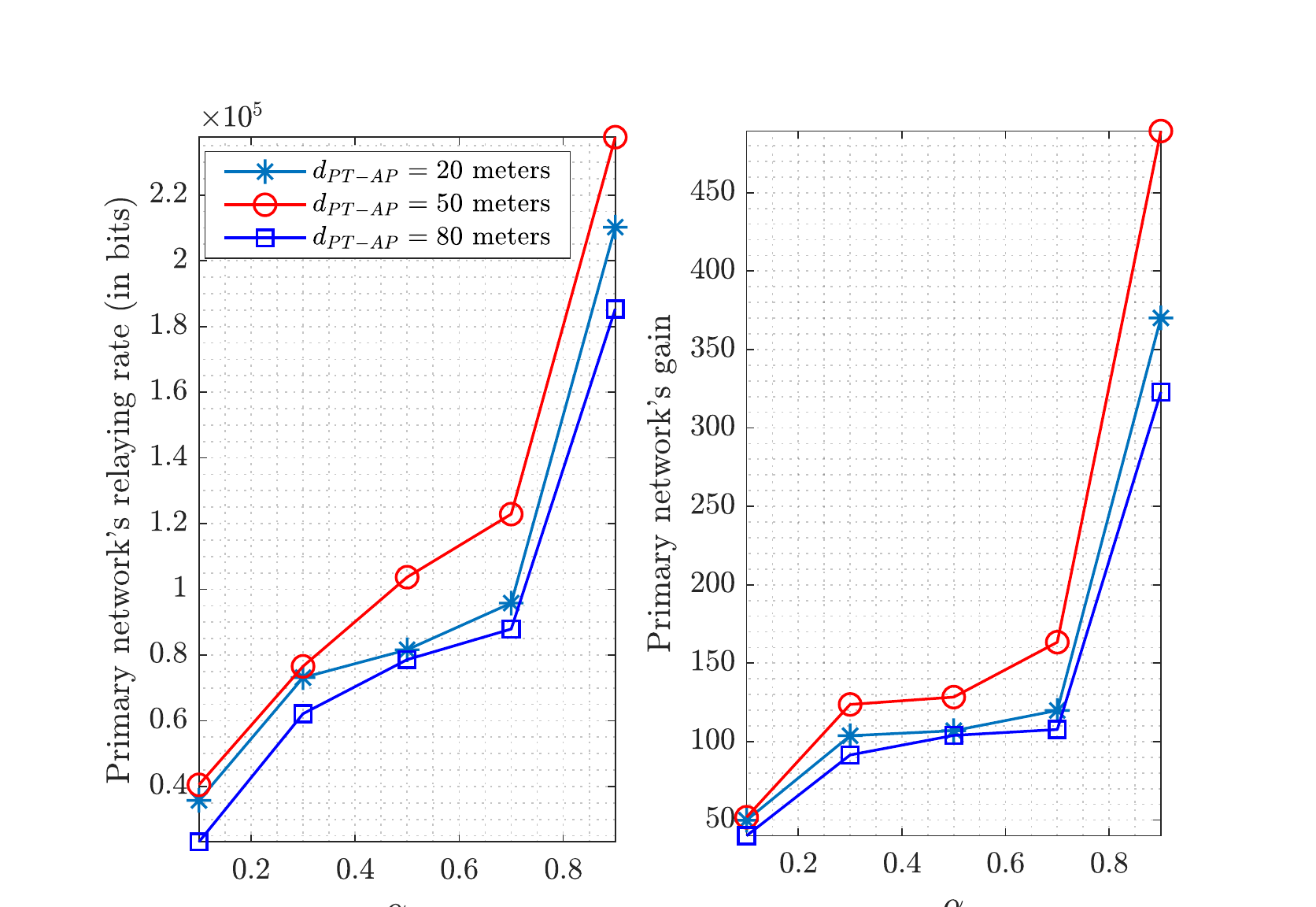}
	\DeclareGraphicsExtensions{.pdf}
	\vspace{-2mm}
	\caption{(a) Relaying rate (b) Primary network gain from spectrum sharing}
	\label{res_fig1}
	\vspace{-6mm}
\end{figure}

In Fig.~\ref{res_fig1}, we plot relaying rate and the primary network's gain from the spectrum sharing for varying $\alpha$, where the primary network's gain is defined as the ratio between relaying rate and direct link rate:
\begin{align}
	R_{\text{relay}}/[ TB_w\log_2(1+\frac{P_{\text{PT}}\left|h_{\text{PT-PR}}\right|^2d_{\text{PT-PR}}^{-\eta_1}}{\sigma^2})]\label{pt_gain}
\end{align}
We consider identical stored energies at IoT nodes, i.e., $E_k = 1$ Joule, $\forall{k}$. Based on the position of the IoT access point, we consider three different scenarios, i.e., $d_{\text{PT-AP}}=20$, 50, and 80 meters, which makes $d_{\text{AP-PR}}=80$, 50, and 20 meters, respectively. We observe that as we increase the value for $\alpha$, relaying rate increases because of higher relaying time duration, i.e., $\tau_r$, which increases relaying rate as can be seen in Fig.~\ref{res_fig1}(a). As the IoT access point moves towards the PR, relaying channel's (i.e., between the IoT access point and the PR) quality improves, which results in better relaying rate, which is observed in Fig~\ref{res_fig1}(a) when the IoT access point is moved from 20 meters to 50 meters from the PT. However, if the IoT access point moves far from the PT, then the channel between the PT and the IoT access point deteriorates significantly, which effectively reduces the value for $\min\left\{\gamma_{\text{PT-AP}},\gamma_{\text{PT-PR+AP-PR}}\right\}$ in Equation~\eqref{eq1} and therefore, relaying rate reduces. This is observed for $d_{\text{PT-AP}}=80$ meters in Fig~\ref{res_fig1}(a). Corresponding gain for the primary network from the cooperation is shown in Fig.~\ref{res_fig1}(b). As relaying rate increases with $\alpha$, from Equation~\eqref{pt_gain}, we can see that the primary network's gain would increase, which can be observed from Fig.~\ref{res_fig1}(b). 

\subsection{IoT network's computation rate and gain for varying $\alpha$}
\begin{figure}[!h]
	\vspace{-2mm}
	\centering
	\includegraphics[trim=0cm 10cm 0cm 8cm,clip=true,width=8cm]{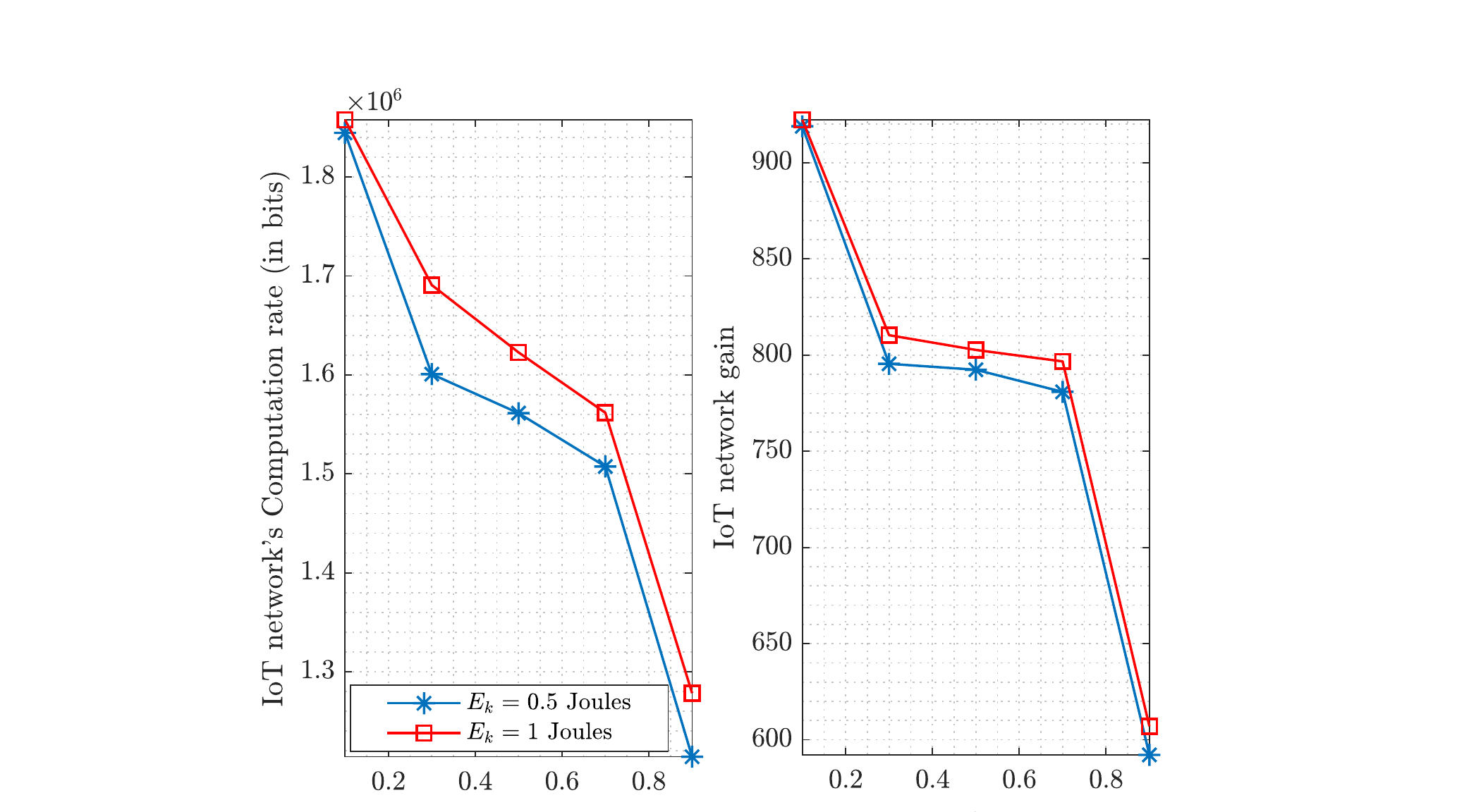}
	\DeclareGraphicsExtensions{.eps}
	\vspace{-2mm}
	\caption{(a) Computation rate (i.e., offloading+local computation) (b) IoT network gain from spectrum sharing}
	\label{res_fig2}
	\vspace{-4mm}
\end{figure}

In Fig.~\ref{res_fig2}, we show computation rate (summation of offloading and local computation rates) and the IoT network's gain from the spectrum sharing. The IoT network's gain is defined as the ratio between total computation rate (summation of offloading and local computation rates) and only local computation rate:
\begin{align}
	 \sum_{k=1}^M[R_{k}+R_{\text{MEC},k}]/[\sum_{k=1}^M\frac{f_{k}}{C_{k}}\min(T,\frac{E_kC_{k}}{w_{k}f_{k}^3})]\label{iot_gain}
\end{align}
We consider $d_{\text{PT-AP}}=70$ meters and $d_{\text{AP-PR}}=30$ meters, and two different set of energy values at IoT nodes, i.e. $E_k=0.5$ Joule and  $E_k=1$ Joule, $\forall{\text{IoT}_k}$. As we increase the value for $\alpha$, relaying time increases, which reduces offloading time duration. In this situation, IoT nodes spend more of their energies on local computation, which increases the total local computation rate. However, we observe that as offloading rate dominates over the local computation rate, effective computation rate reduces in Fig.~\ref{res_fig2}(a). As we increase $\alpha$, the IoT network's offloading rate reduces as more time is spent behind relaying, which effectively reduces the total computation rate for the IoT network, and hence the IoT network's gain from the spectrum sharing as shown in Fig.~\ref{res_fig2}(b).   

\subsection{Comparison with equal processor allocation}
\begin{figure}[!h]
	\centering
	\includegraphics[trim=0cm 10cm 0cm 8cm,clip=true,width=7cm]{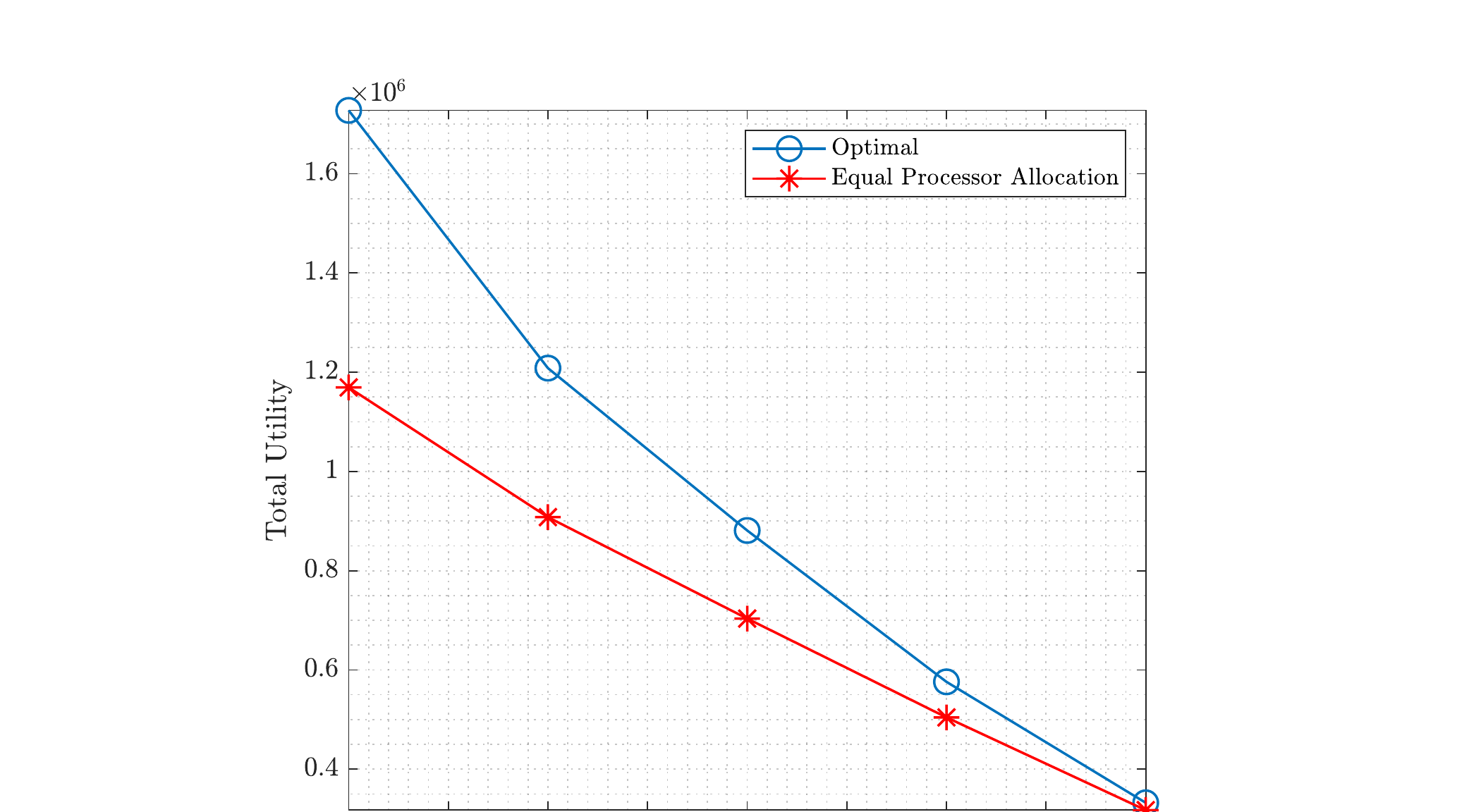}
	\DeclareGraphicsExtensions{.eps}
	\caption{Comparison between our proposed joint optimization and Equal processor allocation}
	\label{res_fig3}	
\end{figure}

To understand the efficacy of joint optimization, i.e., over relaying time, processor allocation, and offloading time and energy allocations, we compare our proposed optimization, i.e., $P1$, with the optimization problem with equal processor allocation, which has been considered in  \cite{liang2019multiuser}. However our system model and system model in \cite{liang2019multiuser} are different. Therefore, we can not directly compare with \cite{liang2019multiuser}. We formulate an equivalent optimization problem for our system model under equal processor allocation at the MEC server, i.e.:
\begin{subequations}
	\begin{align} &\underset{\tau_r,\mathbf{t_{\text{loc}}},\mathbf{t_{\text{off}}},\mathbf{e_{\text{off}}},f_{\text{MEC},k}=f_{\text{MEC}}^{\text{max}}/M}{\text{maximize}}  \alpha R_{\text{relay}}+(1-\alpha)\sum_{k=1}^M[R_{k}+R_{\text{MEC},k}]\nonumber\\
	&\text{subject to:}\hspace{5mm} \text{Equation}~\eqref{eq2-2}, \eqref{eq2-4}, \eqref{eq2-5}, \eqref{eq2-6}, \eqref{eq2-7},  \eqref{eq2-9}.\nonumber
	\end{align}
\end{subequations}
We can solve the above optimization problem following steps for solving $P1$. In Fig.~\ref{res_fig3}, we compare optimal (i.e., $P1$) with above mentioned optimization problem for equal processor allocation, in terms of total utility (i.e., summation of relaying and computation rates) for $d_{\text{PT-AP}}=80$ meters and $d_{\text{AP-PR}}=20$. Moreover, we consider $E_k=1$ Joule, $\forall{k}$. We observe that the result received for the optimal optimization problem, i.e., $P1$, outperforms the result for the optimization problem with equal processor allocation. As we increase the value for $\alpha$, we observe that gap between the results of these two optimization problems reduces. This is due to the difference in computation rate for both optimization methods. For lower values of $\alpha$, the computation rate is given more weight than the relaying rate; therefore, more time is given for offloading, which is judiciously utilized among IoT nodes in the optimal optimization problem $P1$. In equal processor allocation, this advantage is not observed. However, as we increase the value for $\alpha$, relaying rate receives more weight and in that case, the difference between computation rates for these two optimization methods reduces, which reduces the gap between total received utilities for these two optimization problems.

\section{Conclusion}
Both primary and IoT networks get a few hundred times to benefit from the cooperation, which we show in Fig.~\ref{res_fig1}(a) and (b). We observe that our proposed joint optimization method outperforms the equal processor allocation at the MEC server. As in joint optimization, we allocate processors based on IoT nodes' offloading capability, processors are more efficiently utilized. We have considered a centralized optimization problem. Distributed design for such networks may be a relevant future work.

\section*{Acknowledgement}
This work was supported by F.R.S.-FNRS under the EOS
program (EOS project 30452698) and by INNOVIRIS under
the COPINE-IOT project.

\bibliographystyle{References/IEEEtran}
\bibliography{References/references} 

\end{document}